\documentstyle[aps,prb]{revtex}

\begin{document}

\input epsf.sty
\twocolumn[\hsize\textwidth\columnwidth\hsize\csname %
@twocolumnfalse\endcsname

\draft

\widetext

\title{Diagonal static spin correlation in the low temperature 
orthorhombic {\it Pccn} phase of 
La$_{1.55}$Nd$_{0.4}$Sr$_{0.05}$CuO$_{4}$}

\author{S. Wakimoto\footnote{Also at Massachusetts Institute of 
Technology, Cambridge, MA 02139}\footnote{Present address:
Department of Physics, University of Toronto, 
Toronto, Ontario, Canada M5S 1A7} and J. M. Tranquada}
\address{Physics Department, Brookhaven National Laboratory, Upton, New 
York 11973}
\author{T. Ono, K. M. Kojima and S. Uchida.}
\address{Department of Applied Physics, University of Tokyo,
Hongo 7-3-1, Bunkyo, Tokyo 113-8656, JAPAN}
\author{S. -H. Lee and P. M. Gehring}
\address{National Institute of Standards and Technology, NCNR, 
Gaithersburg, Maryland 20889}
\author{R. J. Birgeneau}
\address{Department of Physics and Center for Materials Science and 
Engineering, Massachusetts Institute of Technology, Cambridge, 
Massachusetts 02139 \\and \\
Department of Physics, University of Toronto, 
Toronto, Ontario, Canada M5S 1A7}

\date{\today}
\maketitle

\vspace{-0.1in}

\begin{abstract}

Elastic neutron-scattering measurements have been performed on 
La$_{1.55}$Nd$_{0.4}$Sr$_{0.05}$CuO$_{4}$,
which exhibits a structural phase transition at $T_{s} \sim 60$~K
from the low temperature orthorhombic {\it Bmab} phase 
(labelled LTO1) to 
the low temperature orthorhombic {\it Pccn} phase (labelled LTO2).
At low temperatures, well below $T_{s}$,
elastic magnetic peaks are observed at the ``diagonal'' 
incommensurate (IC) positions $(0, 1\pm0.055, 0)$, with the modulation 
direction only along the orthorhombic $b$-axis just as in
Nd-free La$_{1.95}$Sr$_{0.05}$CuO$_{4}$.  
In the present study, the one-dimensionality of the IC
modulation, which is naturally explained by a stripe model, 
is clearly demonstrated with our ``single-domain" crystal.
The temperature dependence of the IC peak intensity suggests a 
substantial contribution from the Nd$^{3+}$ spins below $\sim3$~K.  
Consistent with this, the $L$ dependence of the magnetic scattering 
is accurately accounted for 
by a model in which the contribution of the Nd$^{3+}$ 
spins is explicitly included.

\end{abstract}


\phantom{.}
]
\narrowtext

%
%
%

\section{Introduction}

%
The relationship between the microscopic 
magnetism and superconductivity is one of the
central issues in the field of high-$T_{C}$ superconductivity.
In particular, La$_{2-x}$(Sr,Ba)$_{x}$CuO$_{4}$ (LSCO, LBCO) and 
related compounds have received intensive attention because of 
their rich magnetic and transport properties.  
In addition, these materials have a simple layered structure with 
single CuO$_{2}$ planes composed of square Cu$^{2+}$ lattices~\cite{M.A.Kastner_98}; 
this facilitates the application of theoretical models.
%
%
It is well known that superconducting LSCO samples exhibit dynamic
incommensurate (IC) magnetic correlations modulated along the direction parallel
to the Cu-O-Cu bonds in the low-temperature orthorhombic (LTO1, $Bmab$)
structure.~\cite{Yoshizawa_88,Bob_89,S.W.Cheong_91}
After the discovery of this IC nature, a systematic neutron-scattering study 
on the superconducting LSCO compounds~\cite{K.Yamada_98} 
revealed a linear relation between the
hole concentration $x$ and the incommensurability parameter $\delta$ (Ref. 6) 
in the under-doped region $(0.06 \leq x \leq 0.12)$, suggesting a strong correlation 
between the superconductivity and the dynamic IC modulation.

%
On the other hand, several investigations have been performed at the 
specific hole concentration $x \sim 1/8$ where for many co-dopants 
the superconductivity is dramatically suppressed.
This so-called 1/8 anomaly was originally discovered in the LBCO 
system~\cite{Moodenbaugh_88} and found to be associated with a structural 
transition to the low-temperature tetragonal (LTT, $P4_{2}/ncm$) phase.~\cite{Axe_89} 
A similar, but much smaller, suppression of $T_{C}$, as well as an enhancement 
of the static magnetic correlations, has been reported in the LSCO system, in 
which there is no transition to 
the LTT phase.~\cite{Kumagai,Goto,T.Suzuki_98,Kimura_99}
%
%
An important clue relevant to the 1/8 anomaly was 
the observation in the 
La$_{1.48}$Nd$_{0.4}$Sr$_{0.12}$CuO$_{4}$ compound of 
very clear elastic magnetic peaks at the {\it parallel} IC 
positions around $(\pi, \pi)$ by neutron-scattering.~\cite{Tra_nature,Tra_prb}
(The substituted Nd$^{3+}$ ions induce the LTT structure as well as the 
1/8 anomaly.~\cite{Crawford_91}  Note that Nd$^{3+}$ ions introduce no 
holes into the system.)
On the basis of the stripe model~\cite{Emery_94} it was suggested that charge
stripes along the Cu-O-Cu bond, which result in {\it parallel} IC
peaks, might be pinned by the corrugation of the CuO$_{2}$ plane 
caused by the coherent tilt
of the CuO$_{6}$ octahedra which is 
perpendicular to the stripes in the LTT structure. 
Thence, the elastic IC
correlations are enhanced and the superconductivity is more suppressed
than in the LTO1 structure of LSCO.

%
Recently Wakimoto {\it et al.}~\cite{waki_rapid,waki_full} discovered the 
so-called ``diagonal" IC peaks in insulating La$_{1.95}$Sr$_{0.05}$CuO$_{4}$,
which shows the LTO1 structure.  
The IC peak positions are shown in Fig. 1(a).  
In this case, the IC modulation is parallel to the orthorhombic 
$b$-axis, which is the same as the coherent tilting direction of the CuO$_{6}$ 
octahedra, and at 45$^{\circ}$ to the Cu-O bonds.
Assuming that the magnetic peaks are associated with charge stripe order, 
the charge stripes would run parallel to the orthorhombic $a$-axis.
Thus, similar to the case of La$_{2-x-y}$Nd$_{y}$Sr$_{x}$CuO$_{4}$ 
(LNSCO), the stripes may be pinned by the 
corrugation of the CuO$_{2}$ planes in the LTO1 phase; 
one might then speculate that the pinning is responsible for 
the insulating behavior below $\sim 100$~K.
Subsequent experiments by Matsuda {\it et al.}~\cite{Matsuda_00} 
and by Fujita {\it et al.}~\cite{Fujita_00}
have shown that this diagonal one-dimensional (1D) spin density wave state in 
LSCO extends across the entire spin-glass region 
$0.02 \stackrel{<}{\sim} x \stackrel{<}{\sim} 0.06$.

\begin{figure}
 \centerline{\epsfxsize=2.5in\epsfbox{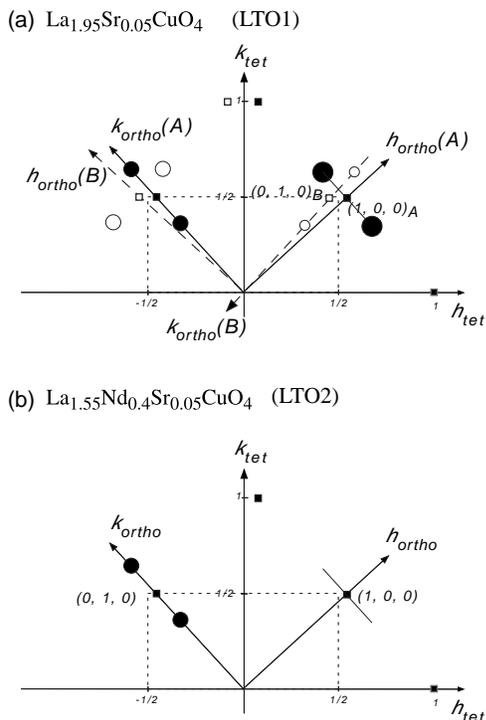}}
  \caption{Incommensurate peak geometry of elastic magnetic peaks 
observed in La$_{1.95}$Sr$_{0.05}$CuO$_{4}$ (upper) and 
La$_{1.55}$Nd$_{0.4}$Sr$_{0.05}$CuO$_{4}$ (lower).
Circles and squares represent the magnetic and the nuclear Bragg peak positions,
respectively.  Note that the Bragg peak positions are observed by $\lambda/2$
neutrons.  In the upper figure, closed and open symbols correspond to
different twin domains.}
\end{figure}

%
These facts naturally suggest the importance of research on the
relation among the three factors: crystal structure, IC magnetic correlations 
and the superconductivity.
Therefore, it is important to investigate how the magnetic and 
transport properties change when the corrugation of the CuO$_{2}$ plane is 
changed in La$_{1.95}$Sr$_{0.05}$CuO$_{4}$.
With the above as motivation, we performed neutron-scattering experiments 
and resistivity measurements 
on La$_{1.55}$Nd$_{0.4}$Sr$_{0.05}$CuO$_{4}$ 
whose corrugation pattern of the CuO$_{2}$ planes should be 
different from that of Nd-free La$_{1.95}$Sr$_{0.05}$CuO$_{4}$. 
Specifically, La$_{1.55}$Nd$_{0.4}$Sr$_{0.05}$CuO$_{4}$ shows the 
low-temperature orthorhombic (LTO2, $Pccn$) structure in which the 
CuO$_{6}$ octahedral tilt direction is slightly rotated from the 
orthorhombic $b$ axis.

%
The format of this paper is as follows.  
The sample preparation, details of the experimental procedure and 
the basic characterizations are 
described in Sec.II.  
The results of elastic neutron-scattering experiments 
are described in Sec.III.
We discuss the relation between the structure and 
magnetic IC modulation in Sec.IV.

\section{Sample preparation and experimental details}

%
Single crystals of La$_{1.55}$Nd$_{0.4}$Sr$_{0.05}$CuO$_{4}$  
were grown by the travelling-solvent floating-zone method.
(Two crystals were prepared in the same manner as described below. 
Most of the data presented here are obtained from one of them, 
and the other one reproduced the magnetic and transport properties.  
Therefore, we do not distinguish two crystals in this paper.)
Dried powders of La$_{2}$O$_{3}$, Nd$_{2}$O$_{3}$, SrCO$_{3}$ 
and CuO of 99.99~\% purity were mixed and baked in air at 
950$^{\circ}$C and 1000$^{\circ}$C for 24 hours with grinding 
between each baking.
The powder sample so-obtained was confirmed to be a single 2-1-4 phase
by X-ray powder diffraction.
Feed rods were shaped in rubber tubing pressed 
by a hydro-static press, and baked in air at 1100$^{\circ}$C for 12 hours.
Solvents with the composition
of La$_{1.55}$Nd$_{0.4}$Sr$_{0.05}$CuO$_{4}$ : CuO = 30 : 70 in molar 
ratio were chosen based on the phase diagram for the pure 
La$_{2}$CuO$_{4}$ compound.~\cite{phasediagram} 
The growth was performed in a four-ellipsoidal-mirror type image furnace 
in an oxygen atmosphere.
The pelletized solvent was placed between the feed rod (attached to the upper 
shaft) and a seed crystal (on the lower shaft), and was melted at the 
focal point.
A crystal was grown by moving the ellipsoidal-mirrors upward.  
During the high temperature operation of growth, there is vaporization 
of a small amount of CuO from the molten zone which causes a change in the
solvent composition.
To avoid this, extra CuO of $\sim 1$~mol\% was added into the feed rods 
to compensate for the loss of CuO vaporizing from the molten zone during the growth. 
Since the concentration of Nd and Sr in the crystallized part
in the beginning of the growth 
is possibly different from the nominal one, 
we continued the growth for more than 100 hours with a growth speed 
of 0.8 mm/hour to achieve the equilibrium condition;
this technique produces a crystal with the nominal composition. 

%
Since we realized that Nd-free LSCO crystals in the lightly doped region,
$0.03 \leq x \leq 0.05$, tend to absorb excess oxygen in the melt-grown 
process in an oxygen atmosphere,~\cite{waki_SG} the as-grown crystal was 
annealed in an Ar atmosphere at 850~$^{\circ}$C for 12 hours to purge 
any excess oxygen.  
%
%
After the treatment, the sample exhibits a spin-glass behavior 
below $\sim5$~K in the magnetic susceptibility measured along an 
arbitrary direction.
The spin-glass transition at $\sim5$~K is consistent with 
that observed in Nd-free La$_{1.95}$Sr$_{0.05}$CuO$_{4}$.~\cite{waki_SG}

%
We also performed resistivity measurements.
The measurements were carried out by the standard dc four-probe method.
Electrodes were attached on the crystal surface with gold-paste.
To achieve good contact between the crystal and the electrodes, the crystal with 
pasted electrodes was annealed at 900~$^{\circ}$C in oxygen for an hour.
Such short time annealing should not affect the oxygen content significantly.
Figure 2 shows the temperature dependence of the in-plane resistivity 
together with that of Nd-free La$_{1.95}$Sr$_{0.05}$CuO$_{4}$. 
The resistivity shows insulating behavior at low temperature, and there is no 
superconducting transition.
There is a small anomaly at the structural transition temperature, $T_{s}$, 
from the LTO1 to the LTO2 phase at $\sim 60$~K,
consistent with that found by neutron-scattering (see Sec. III-A).

\begin{figure}
 \centerline{\epsfxsize=2.5in\epsfbox{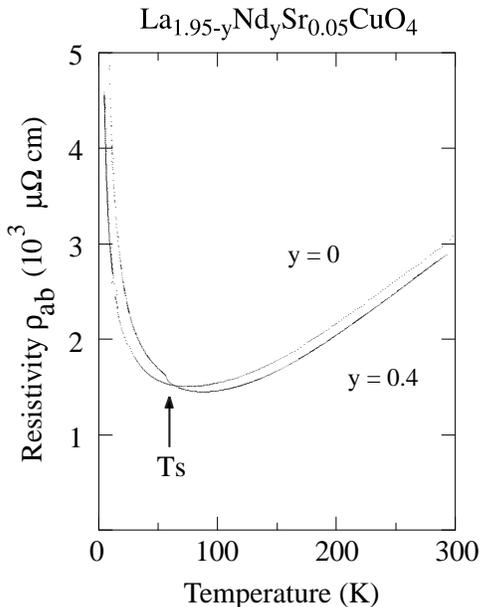}}
  \caption{Temperature dependence of in-plane resistivity 
  together with that for the Nd-free sample.  The temperature indicated 
  as $T_{s}$ is the LTO1-LTO2 structural transition temperature.}
\end{figure}

%
Neutron scattering experiments were performed 
at the triple-axis spectrometer SPINS installed at the cold 
neutron guide at the NIST research reactor.
The horizontal collimator sequence 32$'$-40$'$-S-40$'$-open and 
an incident neutron energy $E_{i}$=5~meV were utilized.
Pyrolytic graphite (002) was used as both monochromator and analyzer.
Contamination from higher-order neutrons was eliminated partially with a 
single Be filter, and completely with two Be filters.
We confirmed that there is no significant multiple scattering around $(\pi, \pi)$
with this incident energy in elastic scattering measurements.~\cite{multi}
A crystal 40~mm in length for the neutron-scattering experiments was cut from the end 
part of the grown crystal. 
The sample was mounted in either the $(HK0)$ or $(0KL)$ scattering plane in 
a pumped-helium cryostat.
%
%
The lattice parameters were $a=5.349$~\AA, $b=5.355$~\AA\ and 
$c=13.012$~\AA\ at $1.5$~K.

%
Figure 1(b) shows the scattering geometry in the (HK0) 
scattering plane of the present LNSCO crystal.  
The squares indicate the apparent nuclear Bragg peak positions determined by  
neutrons with a half wave length ($\lambda/2$), 
while the circles indicate the IC elastic magnetic peak positions.
Properly, there is no nuclear or magnetic peak at the orthorhombic (100) 
or (010) positions.  
The $\lambda / 2$ measurement is used to determine precisely the orientation 
of the IC peaks with respect to the reciprocal lattice (and possible twin 
domains).

The orthorhombic structure typically has two or more twin domains, as 
shown in Fig. 1(a), which indicates the nuclear and magnetic 
peak geometry in the Nd-free $x=0.05$ crystal containing two 
twin-domains.~\cite{waki_full}
However, in the present crystal, one of the domains is so dominant that 
one can treat it as effectively a single domain crystal.
Indeed, only a single pair of the diagonal type IC peaks is clearly observed 
around the $(010)$ position (see Sec. III and IV).

Since the orthorhombic crystallographic axes in both the LTO1 and LTO2 
structures are defined by the diagonals of the distorted squares of the CuO$_{2}$ 
lattice, the orthorhombic $a^{*}$ and $b^{*}$ axes in reciprocal space 
are defined as shown in Fig. 1.
Throughout the present paper, indices based on the orthorhombic $Bmab$ 
or $Pccn$ crystallographic notation are utilized.  
Since clear IC magnetic peaks are observed only around $(010)$, 
we utilized the $(0KL)$ scattering plane to measure the $L$ dependence of 
the IC peak intensity so that the high-intensity peaks lie in the 
scattering plane.

\begin{figure}
 \centerline{\epsfxsize=2.5in\epsfbox{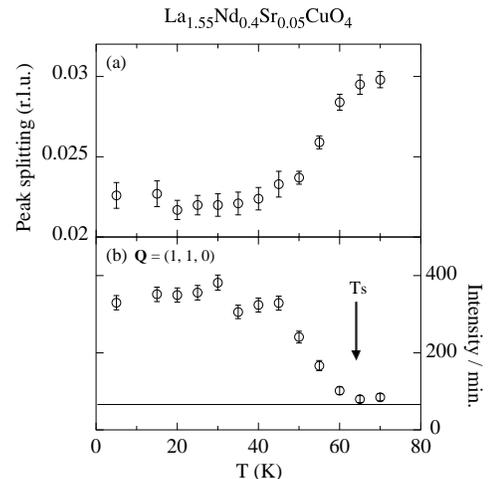}}
  \caption{Temperature dependence of (a) orthorhombic splitting
between (200) and (020), and (b) intensity of superlattice peak
for {\it Pccn} structure at (110).
The solid line is a background level.  $T_{s}$ is the onset of the 
LTO1-LTO2 structural transition.}
\end{figure}

\section{Experimental results}

\subsection{Structural transition to the LTO2 phase}

%
Although one twin domain is dominant in the present crystal, there are 
some other minor twin domains whose volume is estimated to be less than 
1/4 of that of the dominant domain.
Therefore, we were able to observe the orthorhombic splitting between 
the (200) peak of the dominant domain and the (020) peak in a minor domain.
Figure 3(a) shows the temperature dependence of the splitting.
On cooling, the orthorhombicity decreases by about 30~$\%$ between 65~K and 
40~K. 
The intensity of the $(110)$ peak, which is a superlattice peak 
of the $Pccn$ structure but not of the $Bmab$ structure, appears and increases 
in a complementary manner as shown in Fig. 3(b).
These facts demonstrate that the system indeed exhibits a structural 
transition from the $Bmab$ LTO1 phase to the $Pccn$ LTO2 phase.
The transition temperature of $\sim 65$~K agrees well with that expected from the 
phase diagram previously reported by Crawford {\it et al.}~\cite{Crawford_91}
This transition temperature is also in reasonable agreement with that 
determined from the resistivity measurement.

In the LTO1 structure above 65~K, the CuO$_{6}$ octahedra tilt along the 
orthorhombic $b$-axis.
In the LTO2 structure, the octahedra tilt along a direction rotated within 
the plane away from the $b$-axis.
From the change of the orthorhombic splitting in Fig. 3, the shift of the 
tilt direction from the $b$-axis is estimated to be $\sim15$~degrees.

\begin{figure}
 \centerline{\epsfxsize=2.5in\epsfbox{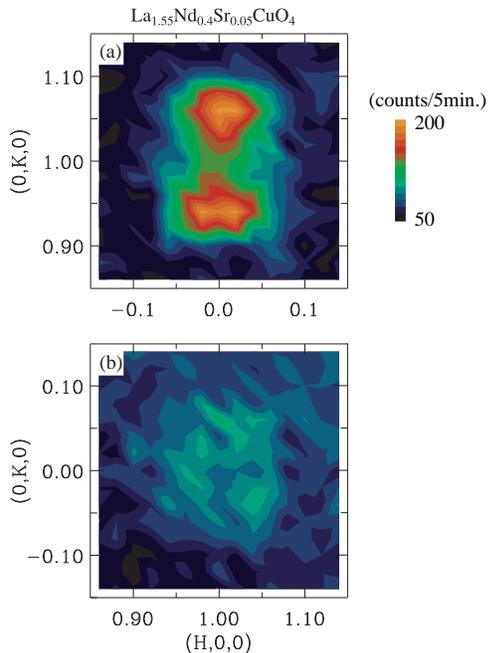}}
  \caption{Contour plots of magnetic peak intensity around 
(a) (010) and (b) (100).  Incommensurate peak positions are summarized
in Fig. 1(b) by closed squares.}
\end{figure}

\subsection{Magnetic cross section}

%
Contour plots of the elastic neutron-scattering intensity around $(010)$ 
and $(100)$ are shown in Figs. 4(a) and 4(b), respectively. 
%
%
Since the crystal has a single dominant twin domain, the 1D 
nature of the IC modulation along the orthorhombic $b$-axis is clearly  
observable in Fig. 4(a);
it is consistent with, but more obvious than, that 
first reported for the Nd-free $x=0.05$ sample.~\cite{waki_full} 
An important difference from the Nd-free $x=0.05$ compound is that 
clear magnetic peaks appear only around the $(010)$ position;
for the Nd-free sample, the intensity is strongest for the peaks split 
about (100).
(The two cases are schematically summarized in Fig. 1.)
A similar change of the magnetic peak position from (100) to (010) 
has been previously reported in Sr-free 
La$_{2-y}$Nd$_{y}$CuO$_{4}$.~\cite{Crawford_93,Keimer_Nd}
This feature is discussed in Sec.IV in terms of the spin orientation.

%
To analyze the IC peaks in detail, we made scans 
through the peak positions in the vicinity of $(010)$ with higher statistics,
achieved by optimizing the vertical focusing of the incident 
neutron beam.
Figures 5(a) and 5(b) show the peak profiles along the trajectories 
$\alpha$ and $\beta$ indicated by arrows in Fig. 5(d).
The incommensurate peaks are observed at the $(0, 1 \pm \epsilon, 0)$ 
positions.
(The IC peak intensity in unit time is different from that in Fig. 4 
due to the change in the vertical focusing of the incident beam;  
however, the vertical focusing does not affect the intrinsic 
characteristic values, such as the width, positions and temperature 
dependence of the IC peaks.)

\begin{figure}
 \centerline{\epsfxsize=2.5in\epsfbox{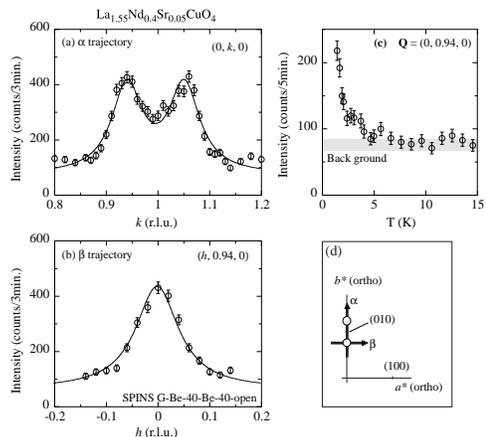}}
  \caption{Peak profiles along (a) the $\alpha$ trajectory and
(b) the $\beta$ trajectory shown in (d).  Solid lines are
fitting results by Lorentzian function.  By this fitting,
the incommensurability parameter $\epsilon$ is $\sim 0.055$.
(c) Temperature dependence of elastic IC magnetic peak 
at $(0, 0.94, 0)$.}
\end{figure}

%
The solid lines in Fig. 5 are fitted curves corresponding to a 
two-dimensional Lorentzian function 
convoluted with the instrumental resolution.
The intrinsic peak half widths along the $a^{*}$ and $b^{*}$ directions are
$\kappa_{a}=0.053$~\AA$^{-1}$ and $\kappa_{b}=0.039$~\AA$^{-1}$, respectively.
The incommensurability parameter is $\epsilon=0.055 (\pm 0.004)$, which is slightly
lower than that for Nd-free $x=0.05$, where $\epsilon=0.064$.~\cite{waki_full} 
%
%
Although in the contour plot of Fig. 4(b) the peak at $(0, 1-\epsilon, 0)$ 
seems to be split additionally along the $a^{*}$-axis, the profile in 
Fig. 5(b) with higher statistics shows a single peak centered at $h=0$.
%


%
The temperature dependence of the IC peak at (0, 0.94, 0) is shown in 
Fig. 5(c).
This measurement was done without optimizing the vertical focusing.
The intensity gradually increases with decreasing temperature below 
$\sim 5$~K and rapidly increases below $\sim 3$~K.
From previous experience with Nd-substituted samples,~\cite{Tra_prb}
we expect that this additional increase of intensity below $\sim 3$~K is caused 
by an additional ordering of the Nd$^{3+}$ spins.

\begin{figure}
 \centerline{\epsfxsize=2.5in\epsfbox{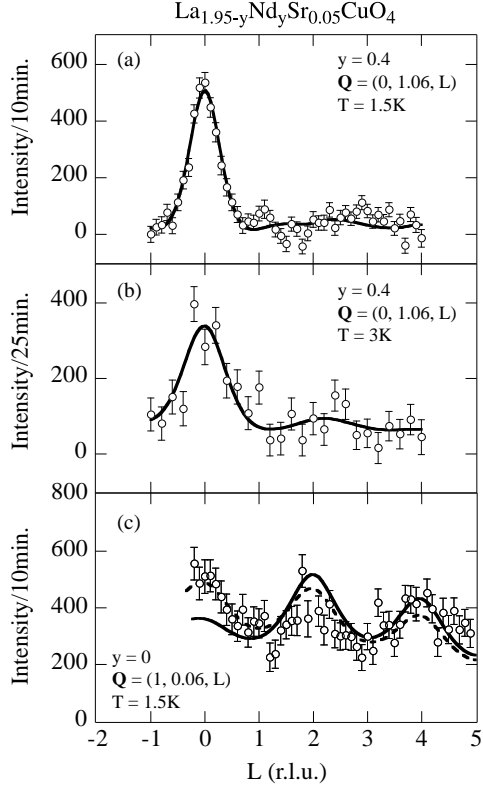}}
  \caption{$L$ dependence of the IC peak intensity in 
  (a) La$_{1.55}$Nd$_{0.4}$Sr$_{0.05}$CuO$_{4}$ at 1.5K, 
  (b) La$_{1.55}$Nd$_{0.4}$Sr$_{0.05}$CuO$_{4}$ at 3K, and (c) Nd-free 
  La$_{1.95}$Sr$_{0.05}$CuO$_{4}$ at 1.5K (obtained from Ref. 18).  
  The back ground intensity at 30K have been subtracted.  
  Solid lines in (a) and (b) are the results of fit by a model including 
  contribution from Nd$^{3+}$ spins and out-of-plane component of Cu$^{2+}$ 
  spins. (See text).  Solid line in (c) is the fit with the same model without 
  Nd$^{3+}$ spin contribution and out-of-plane component of Cu$^{2+}$, 
  while dashed line is the fit without only Nd$^{3+}$ spin component.}
\end{figure}

%
The $L$ dependences of the intensity at the IC position $(0, 0.94, L)$ at 1.5~K and 3~K 
are shown in Figs. 6(a) and 6(b), respectively.
For both results, the intensity at 30~K was subtracted as background so 
that the intensities shown are purely magnetic.
%
%
We fit the data with a model introduced previously in a study of the 
Nd-substituted $x=0.12$ compound.~\cite{Tra_prb}
The model has the form:

\begin{equation}
I \propto |F|^{2}
\frac{1-t^{2}}{1+t^{2}-2t \cos{\phi}}.
\label{eq_1}
\end{equation}

This function consists of the magnetic structure factor $F$ and 
an oscillating function with line width determined by 
$t = {\rm exp}(-c / \xi)$, where $\xi$ is a correlation length.
In the analysis of the Nd-substituted $x=0.12$ results, 
an oscillation period of $L=1$ was utilized.
This was an appropriate choice in that case because of the 
assumed rotation of the stripe orientation by 90$^{\circ}$ from 
one CuO$_{2}$ plane to the nearest neighbor plane, which gives rise to 
correlations between the next nearest neighbor planes.
However, in the present system with $x=0.05$ the IC modulation is only 
along the orthorhombic $b$-axis, that is, a 90$^{\circ}$ rotation of the stripe 
orientation is not likely. 
Thus, we utilized the oscillation function with a period of $L=2$; that is,
$\phi=\pi L$.

The structure factor is described as

\begin{equation}
|F|^{2} = p_{\rm Cu, \parallel}^{2} f_{\rm Cu}^{2} + 
[p_{\rm Cu, \perp} f_{\rm Cu} + 
y p_{\rm Nd} f_{\rm Nd} \cos{(2 \pi L z_{\rm Nd})}]^{2}.
\label{eq_2}
\end{equation}
In this equation, 
$f_{\rm Cu}$ and $f_{\rm Nd}$ are {\bf Q}-dependent magnetic form 
factors for the Cu$^{2+}$ and Nd$^{3+}$ spins which are taken from the 
literature.~\cite{Shamoto,Blume}
$z_{Nd}$ is the distance between Nd and the nearest Cu in units of $c$, 
which has been determined to be 0.36 by a neutron powder diffraction 
study.~\cite{Axe_z}
$p$ is the spin component perpendicular to the scattering 
vector {\bf Q}, which relates to the ordered magnetic moment $\mu$
by
${\bf p} = {\rm \bf \vec\mu} - {\rm \bf \hat Q} ({\rm \bf \hat Q} \cdot {\rm \bf 
\vec\mu})$. 
The indices $\parallel$ and $\perp$ represent in-plane and out-of-plane 
components of the Cu$^{2+}$ spin, respectively. 

In this model the Nd$^{3+}$ spins are assumed to be parallel to the $c$-axis 
which is expected to be the easy axis of the Nd$^{3+}$ spins based on 
the magnetic susceptibility measurements for 
La$_{1.3}$Nd$_{0.6}$Sr$_{0.1}$CuO$_{4}$.~\cite{Xu_00}
Therefore, 
the first term of Eq. (\ref{eq_2}) describes the contribution of the in-plane component of 
the Cu$^{2+}$ spins while the second term describes contributions of the 
out-of-plane components of both the Cu$^{2+}$ and Nd$^{3+}$ spins.
We assumed the spin direction of ${\bf p}_{\parallel}$ to be random because 
the system shows the features of a spin-glass.
The fitting results are shown by the solid lines in Figs. 6(a) and 6(b), which agree 
with the experimental results.

\begin{table}
  \caption{Parameters obtained in fitting to the $L$ dependences by Eq. (1). 
$\mu$ is ordered magnetic moment which relates to $p$ in Eq. (2) by 
${\bf p} = {\rm \bf \vec\mu} - {\rm \bf \hat Q} ({\rm \bf \hat Q} \cdot {\rm \bf 
\vec\mu})$.  The values listed at the bottom are given by the fitting for 
the $y=0$ sample data with  $p_{\rm Cu, \perp}$ as a fitting valuable.}
\begin{tabular}[htb]{lccc}
$y$ and $T$(K)
&
$\mu_{\rm Cu, \perp} / \mu_{\rm Cu, \parallel}$
&
$\mu_{\rm Nd} / \mu_{\rm Cu}$
&
$\xi / c$ \\ 
\hline
$y=0.4$,\ $T=1.5$
&
$0.96 \pm 0.25$
&
$4.1 \pm 0.9$
&
$0.58 \pm 0.11$ \\
$y=0.4$,\ $T=3$
&
$1.04 \pm 0.37$
&
$1.1 \pm 0.6$
&
$0.36 \pm 0.18$ \\
$y=0$,\ $T=1.5$
&
0
&
0
&
$0.48 \pm 0.05$ \\
&
($0.55 \pm 0.05$)
&
(0)
&
($0.44 \pm 0.04$) \\
\end{tabular}
\end{table}

%
In order to check the consistency of this function, we also fit the 
$L$ dependence of the IC peak intensity for the Nd-free $x=0.05$ compound 
reported in Ref. 18 with the 
function presented above without the Nd$^{3+}$ spin contribution.
Since the out-of-plane component of the Cu$^{2+}$ spins $p_{\rm Cu, \perp}$
is assumed to be driven by the interaction with the Nd$^{3+}$ spins in the 
present model, we fit the data with $p_{\rm Cu, \perp}$ fixed at zero.
The fitted curve is shown in Fig. 6(c) by the solid line.
For comparison, we also fit the data with $p_{\rm Cu, \perp}$ as a fitting 
variable, as shown by the dashed line.~\cite{fit_exp} 

The parameters obtained by the fitting illustrated in Fig. 6 are summarized in Table 1.
The values listed in the bottom row are obtained by fitting the Nd-free 
data with a $p_{\rm Cu, \perp}$ component.
In the Nd-substituted sample, the out-of-plane component of the Cu$^{2+}$ spins is 
larger than that in the Nd-free sample even if we assume the Nd-free sample 
also has an out-of-plane component.
This indicates that the the out-of-plane component of the Cu$^{2+}$ spins 
is induced by the correlation with the Nd$^{3+}$ spins parallel 
to the $c$-axis.
For the Nd-substituted sample, the ratio $\mu_{\rm Nd} / \mu_{\rm Cu}$ increases 
with decreasing temperature, consistent with the rapid increase of the IC 
intensity below 3~K in Fig. 5(c).

\section{Discussion}

%
The present paper reports the IC magnetic order observed by neutron-scattering 
experiments for the La$_{1.55}$Nd$_{0.4}$Sr$_{0.05}$CuO$_{4}$ 
compound.
%
%
One of the important results is that the same type of 1D IC 
spin modulation as that reported for the Nd-free $x=0.05$ 
compound~\cite{waki_full} is clearly observed in the almost single twin-domain 
sample of Nd-substituted $x=0.05$.
%
%
This demonstrates that the 1D IC modulation along the orthorhombic $b$-axis 
is common in the lightly Sr-doped spin-glass systems with both LTO1 and 
LTO2 structure.

%
A major difference from the Nd-free sample is that the clear IC peaks appear 
around the $(010)$ position in the present LTO2 compound, 
while a more intense pair of peaks appears around $(100)$ in the 
Nd-free $x=0.05$.
A similar change of the magnetic peak position from $(100)$ to $(010)$ 
has been reported by powder neutron-scattering for 
La$_{1.8}$Nd$_{0.2}$CuO$_{4}$;~\cite{Crawford_93} 
the magnetic commensurate peak at $(100)$ in the LTO1 phase shifts to 
$(010)$ below the LTO1-LTO2 transition temperature.
%
%
Such a change of the magnetic peak position 
can be explained simply by a rotation of the spin direction that causes 
a change of the antiferromagnetic propagation vector from 
the $\hat a$ to the $\hat b$ direction.
This change is schematically drawn in Fig. 7.
The square represents the CuO$_{2}$ square lattice. 
The arrows at the corners are spins of the Cu$^{2+}$ ions 
on the same CuO$_{2}$ plane while the arrows at the 
center are spins on the nearest neighbor plane.

\begin{figure}
 \centerline{\epsfxsize=2.5in\epsfbox{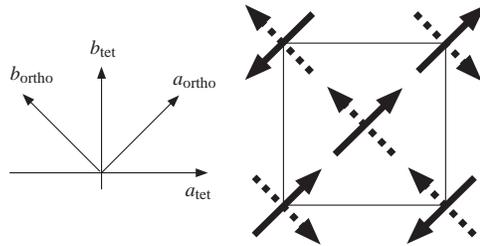}}
  \caption{Schematic figure of the spin structure.  The square represents 
  CuO$_{2}$ square lattice.  The arrows at the corners are spins of the 
  Cu$^{2+}$ ions on the same plane and the arrows at the center are spins 
  on the nearest neighbor plane.  The spin structure shown by the dashed and solid 
  arrows give the magnetic bragg peaks at $(100)$ and $(010)$, 
  respectively.}
\end{figure}

The spin structure shown by the dashed arrows 
has the propagation vector along $\hat a$ 
and gives a magnetic peak at $(100)$ .
On the other hand, the structure shown by the solid arrows 
has the propagation vector along $\hat b$ and accordingly gives the peak at $(010)$ . 
The latter spin orientation, with the modulation along the $b$-axis, 
is consistent with the strong IC peak intensity around $(010)$ 
as observed in the present LTO2 compound.
The change between the two magnetic structures is realized by the opposite rotation 
of the spin directions in neighboring planes; that is, 
the spins on one plane rotate clockwise uniformly while those in the 
nearest-neighbor plane rotate counter-clockwise.
Such an alternative rotation across the LTO1-LTO2 transition temperature has 
also been identified in a neutron-scattering study of La$_{1.65}$Nd$_{0.35}$CuO$_{4}$ 
by comparing the peak intensity at $L=0$ and $L=1$.~\cite{Keimer_Nd}

Another difference between Nd-substituted and Nd-free 
samples can be seen in the IC peak width.
From fitting with the same two-dimensional Lorentzian function convoluted with the 
instrumental resolution, 
the intrinsic widths for the present crystal are determined to be
$\kappa_{a}=0.053$~\AA$^{-1}$ and $\kappa_{b}=0.039$~\AA$^{-1}$,  
which are slightly larger than those for the Nd-free $x=0.05$ sample, 
$\kappa_{a} \sim 0.04$~\AA$^{-1}$ and $\kappa_{b} \sim 0.03$~\AA$^{-1}$. 
~\cite{waki_full}
In order to check whether this larger width is intrinsic to the LTO2 
structure, a more systematic comparison between the LTO2 and LTO1 
phases is required.


%
Finally, we discuss the stripe model.
Among the theoretical explanations for the IC feature,
the 1D modulation observed in the present study is most easily understood 
in terms of the stripe model~\cite{Emery_94}. 
Based on the stripe model, an IC peak profile can be described by the 
same function as Eq.~(\ref{eq_1}).
In this particular case, $\phi=2\pi n k$, 
$t = (-1)^{n+1} \exp{(-nb / \xi)}$ where $n$ is the stripe spacing 
in the orthorhombic lattice unit,
and $F$ is the magnetic structure factor for a single antiferromagnetic region 
between nearest charge stripes.~\cite{Tra_condmat} 
This function gives the oscillation modified by the antiferromagnetic 
structure factor.

\begin{figure}
 \centerline{\epsfxsize=2.5in\epsfbox{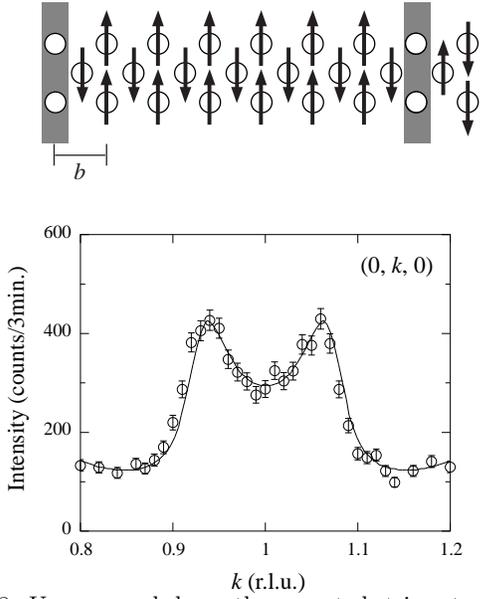}}
  \caption{Upper panel shows the expected stripe structure in the present 
  LNSCO sample.  White circles and arrows represent Cu atoms and their 
  spins.  The shaded regions are charge stripes.  The lower panel is the 
  IC peak profile with the fitted curve using the stripe model. (See text.)}
\end{figure}

In the present system, the charge stripes are parallel to the orthorhombic 
$a$-axis and have the spacing $nb$ 
as schematically shown in the upper panel of Fig. 8.
(This figure corresponds to the spacing $n=7$.)
Therefore, the magnetic structure factor $F$ can be described 
approximately as a function of $k$:

\begin{equation}
F \propto 1+2\sum_{j=1}^{n-1} (-1)^{j}\cos{(\pi jk)}.
\label{eq_4}
\end{equation}
The actual fitted curve using Eqs.~(\ref{eq_1}) and (\ref{eq_4}) with fixed $n=7$ 
is indicated by the solid line in Fig. 8.
The only fitting parameter, $\xi$, is $\sim 31$~\AA.  
(The stripe spacing $n=7$ is smaller than the expected value 
$n=8$ from the incommensurability $\epsilon \sim 0.06$~reciprocal lattice 
unit (r.l.u.). 
However, in the fitting with $n=8$, the damping by the magnetic structure 
factor is stronger than that for $n=7$ and results in poorer agreement with 
the data.)
Figure 8 demonstrates that 
this model gives a reasonable description of the experimental data. 
As noted in section III, in the comparison of the incommensurability parameter 
$\epsilon$ determined by the Lorentzian function fit, 
the Nd-substituted sample has the incommensurability, 
$\epsilon = 0.055$, which is slightly smaller than that for Nd-free sample, 
$\epsilon = 0.064$.
However, in the fitting using the stripe function, 
the fit with the fixed parameter $n=7$ also agrees reasonably with the 
experimental data for Nd-free $x=0.05$ with the fitting variable $\xi \sim 
47$~\AA.
Thus the stripe structure can explain the IC peaks also in the lightly 
hole doped region with a stripe spacing of $7b$ for the $x=0.05$ compounds.

As a future experiment,
it would be interesting to see whether an in-plane anisotropy of
the conductivity, associated with the 1D magnetic modulation,
can be observed in an untwinned crystal with $x \leq 0.05$,
where a unique orientation of the IC modulation has been observed.

\section{Acknowledgement}

We thank Y. Endoh, K. Hirota, and 
G. Shirane for invaluable discussions. 
The work at Brookhaven National Laboratory was carried out under 
Contract No.\ DE-AC02-98CH10886, Division of Materials Sciences, U.\ S. 
Department of Energy. 
The work at MIT was supported by the NSF under Grant No.\ DMR0071256 
and by the MRSEC Program of the National Science Foundation under 
Award No.\ DMR98-08941. 
The work at the University of Tokyo was financially supported by 
Grant-in-Aid for Scientific Research from the Ministry of Education,
Science, Sports and Culture of Japan.
The work at SPINS in National Institute of Standards and Technology is based 
upon activities supported by the National Science 
Foundation under Agreement No. DMR-9986442.
Work at the University of Toronto is part of the Canadian Institute for 
Advanced Research and is supported by the Natural Science and Engineering 
Research Council of Canada.

\vspace{5mm}
\noindent
$\star$ Also at Massachusetts Institute of Technology, Cambridge, MA 02139.

\noindent
$\dagger$ Present address: Department of Physics, University of Toronto, Toronto, 
Ontario, Canada M5S 1A7.


\begin{references}
\bibitem{multi} This confirmation can be done by counting the intensity at $(\pi, \pi)$ 
changing the size of Ewald sphere in reciprocal space; that is, changing 
both incident and final energy so as to always satisfy elastic condition.
\bibitem{fit_exp} Although, from the basic assumption in the model of Eq. (2), the 
Nd-free sample is supposed to have no $p_{\rm Cu, \perp}$ component, the 
fitting with the finite value of $p_{\rm Cu, \perp}$ shown by the dashed 
line in Fig. 6(c) seems to show better agreement with the experimental data.
However, we should note that the contamination at $L=-0.5$ as noted in 
Ref. 18 may cause a discrepancy between the data and the solid line fitted 
with $p_{\rm Cu, \perp}=0$.
\end{references}
\end{document}